\documentclass[twocolumn,aps,pre]{revtex4}
\usepackage{graphicx,color}
\newcommand{\figlml}{%
\begin{figure}[htbp]
   \includegraphics[width=3.5in,clip]{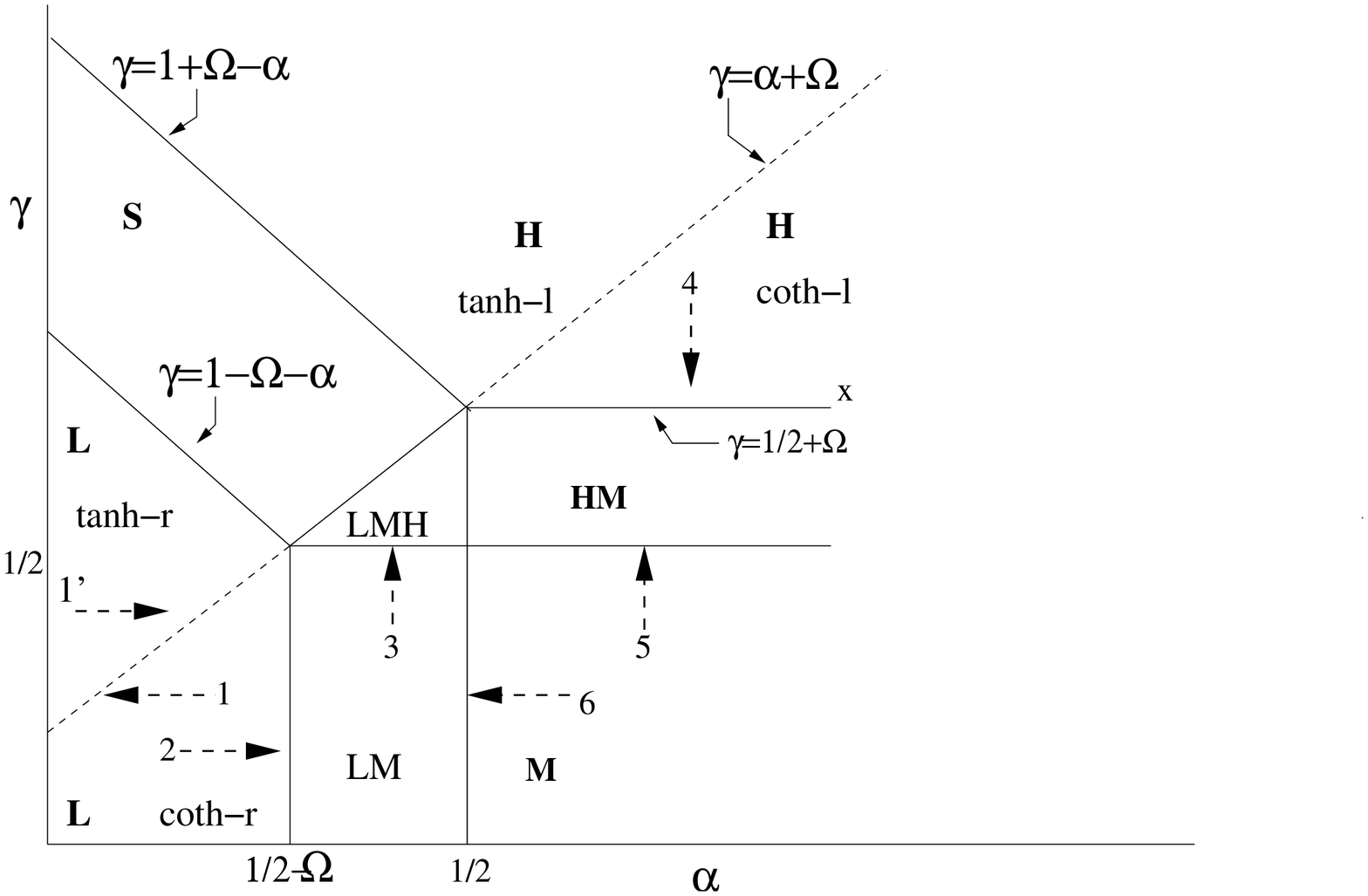}
    \caption{Phase diagram for $K=\omega_a/\omega_d=1$ and 
$\Omega=\omega_d N<.5$. $N$ is the number of lattice points.  
 The coexistence of  a maximal-current and  a low-
 or a high-density phase 
is represented by LM or HM respectively. The coexistence of  
low-density,  maximal-current and  high-density phases is 
similarly represented 
by LMH. Approach to different phase boundaries are 
indicated by the paths
with arrows."l" and "r" indicate presence of boundary layers at left 
and right boundaries respectively. Dashed lines represent surface transition 
lines.}
\label{fig:lml}
\end{figure}
}
\newcommand{\figtanh}{%
\begin{figure}[htbp]
\includegraphics[width=2in,clip]{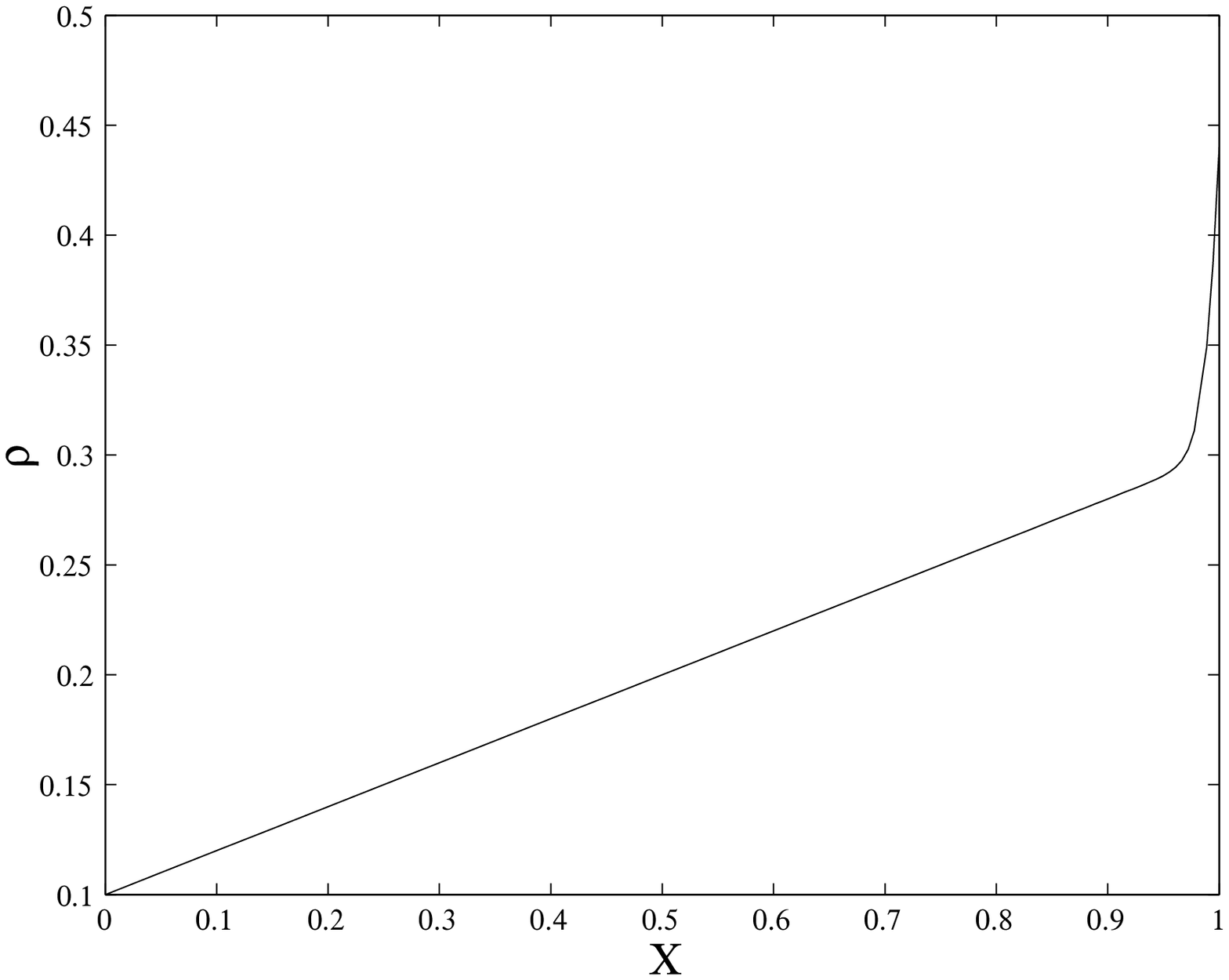}
\caption{Density profile with positive slope at the right boundary 
for $K=1$ with $\alpha=.1$, $\gamma=.44$, $\epsilon=.0035$ and $\Omega=.2$.}
\label{fig:tanh}
\end{figure}
}
\newcommand{\figintersect}{%
\begin{figure}[htbp]
\includegraphics[width=2in,clip]{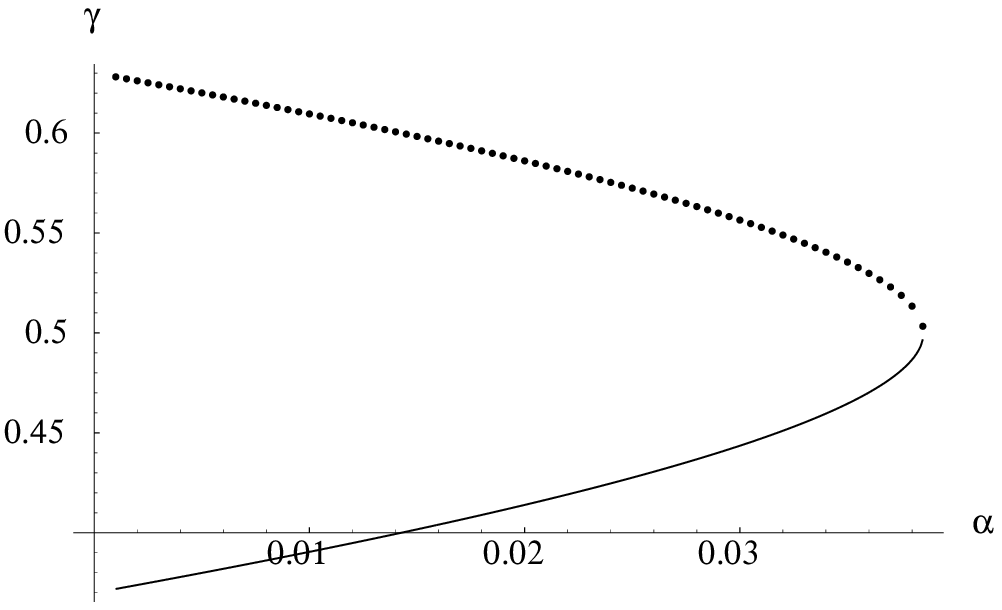}
\caption{Solid line represents the surface transition line in the 
low-density phase for $K=3$ and 
$\Omega=.1$. 
The dotted line corresponds to the phase boundary between the low-density 
and the shock phase.
The  intersection of the two lines is  the critical point 
$(\alpha_c,\gamma_c)$.}
\label{fig:intersect}
\end{figure}
}
\newcommand{\figcoth}{%
\begin{figure}[htbp]
\includegraphics[width=2in,clip]{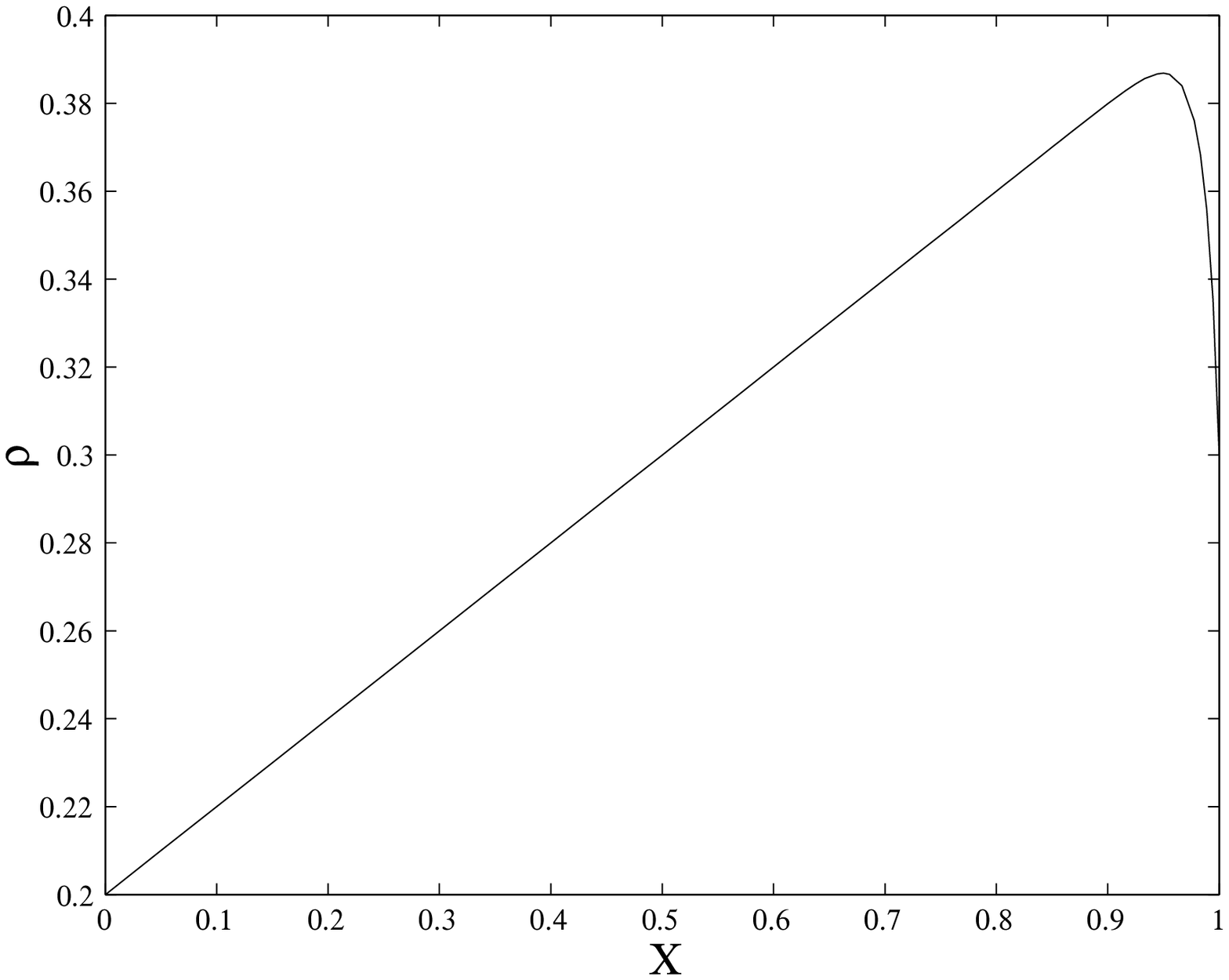}
\caption{Density profile with negative slope at the right boundary 
for $K=1$ with $\alpha=.2$, $\gamma=.3$, $\epsilon=.0035$ and
$\Omega=.2$.}
\label{fig:coth}
\end{figure}
}
\newcommand{\figtanhhigh}{%
\begin{figure}[htbp]
\includegraphics[width=2in,clip]{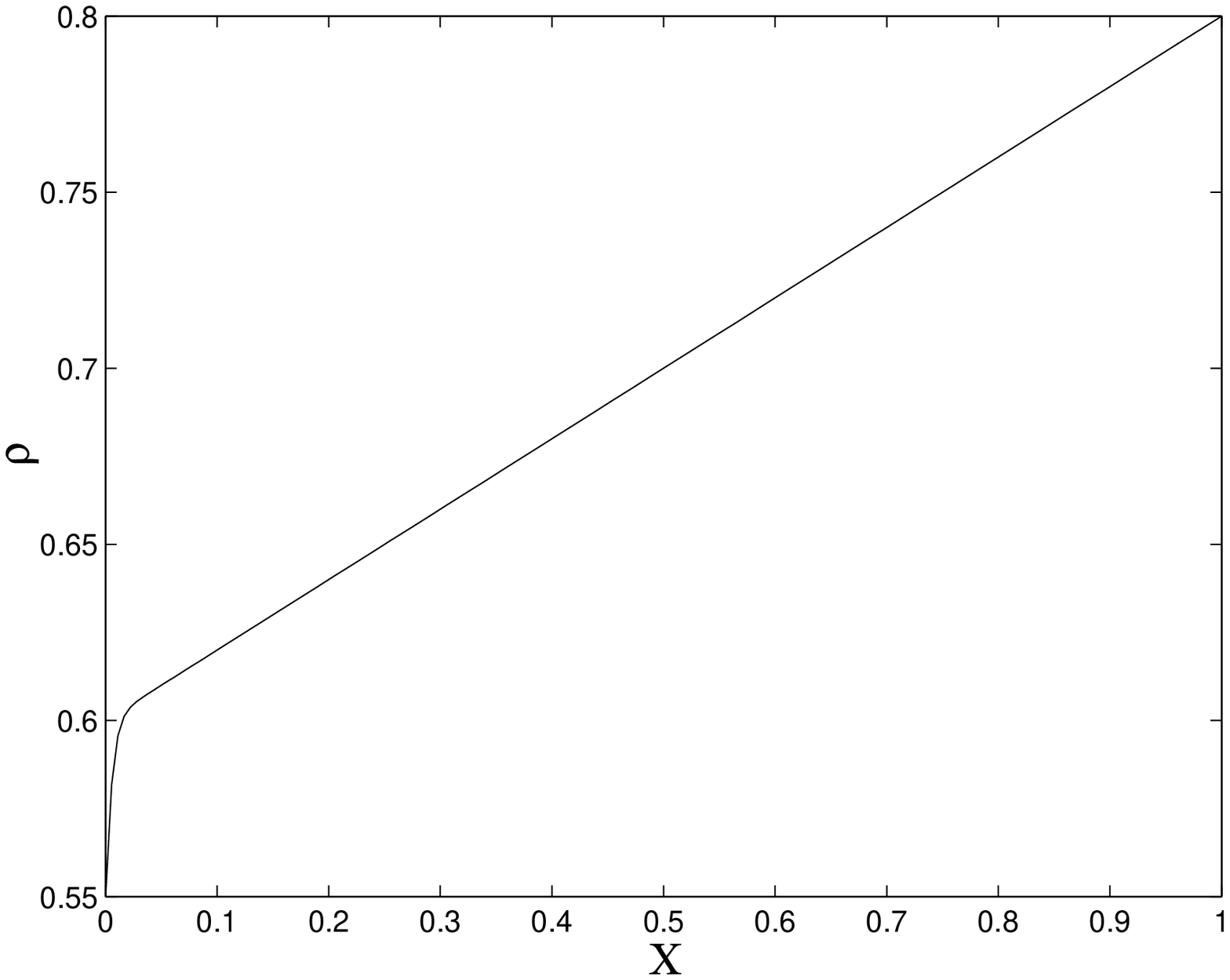}
\caption{Density profile with positive slope at the left boundary 
for $K=1$ with $\alpha=.55$, $\gamma=.8$, $\epsilon=.001$ and $\Omega=.2$.}
\label{fig:tanh-high}
\end{figure}
}
\newcommand{\figcothhigh}{%
\begin{figure}[htbp]
\includegraphics[width=2in,clip]{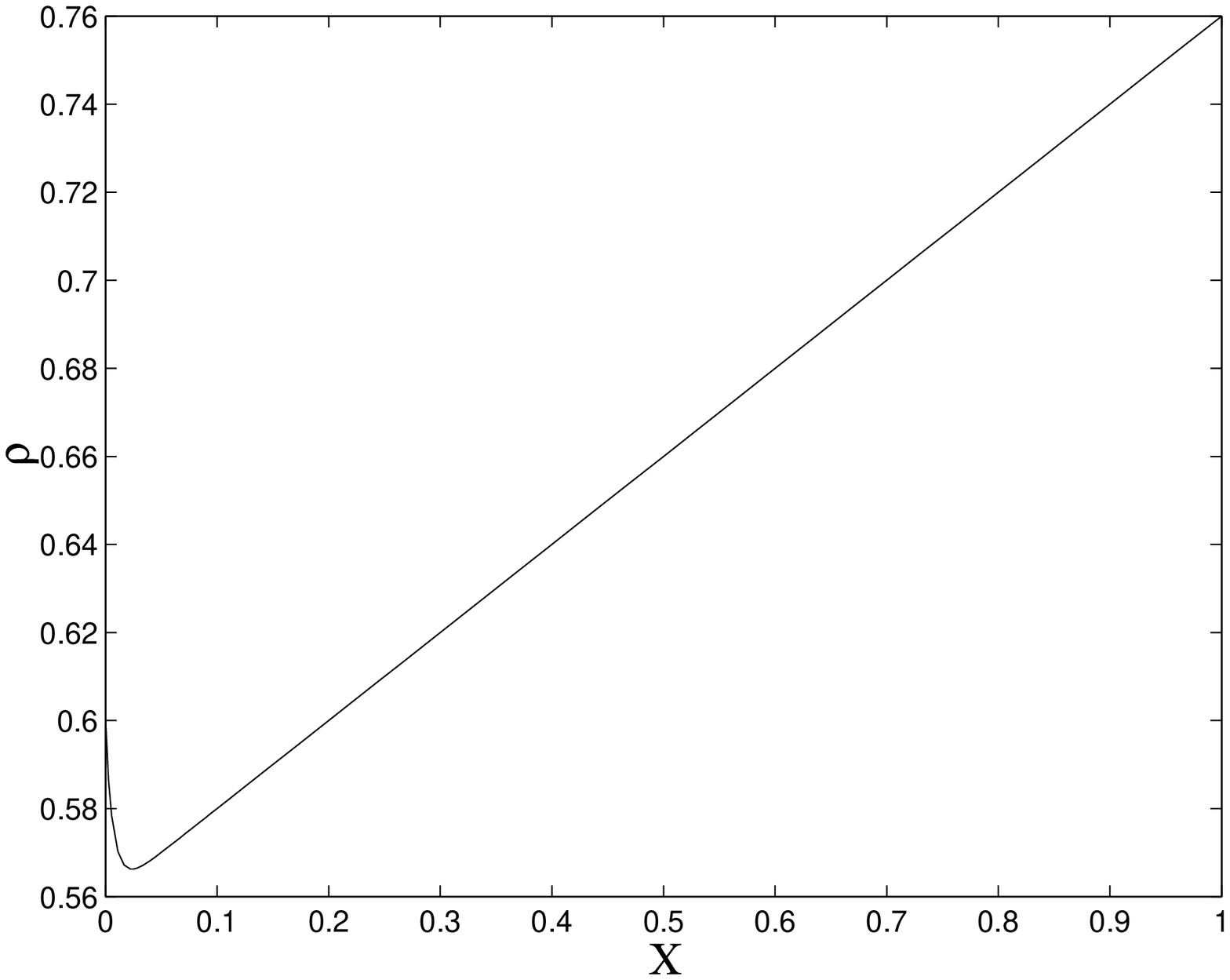}
\caption{Density profile with negative slope at the left boundary 
for $K=1$ with $\alpha=.6$, $\gamma=.76$, $\epsilon=.001$ and $\Omega=.2$.}
\label{fig:coth-high}
\end{figure}
}
\newcommand{\figlm}{%
\begin{figure}[htbp]
\includegraphics[width=2in,clip]{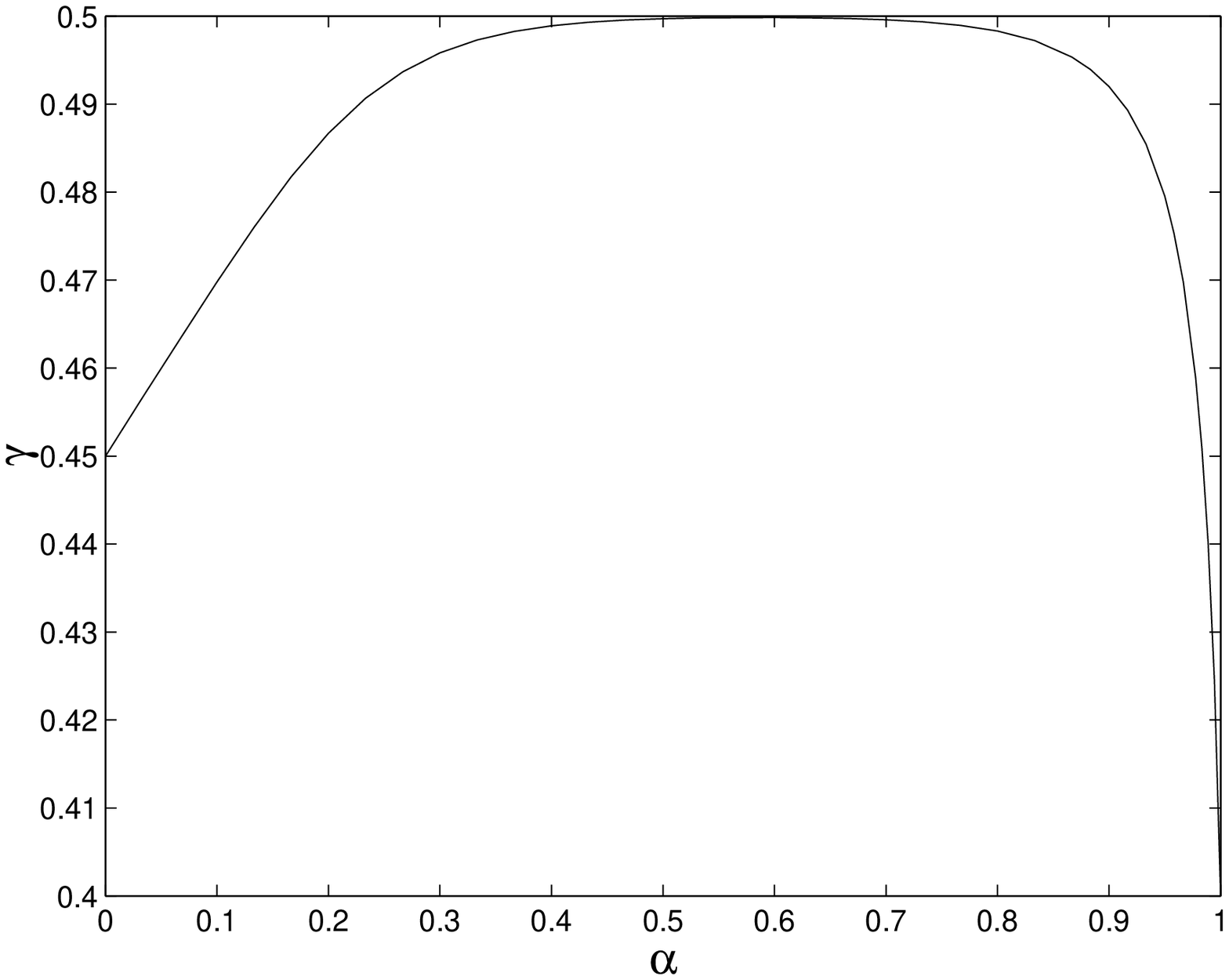}
\caption{Density profile in the low-density maximal current (LM) phase 
with $\alpha=.45$, $\gamma=.4$, $\epsilon=.002$ and $\Omega=.2$.}
\label{fig:lm}
\end{figure}
}
\newcommand{\fighm}{%
\begin{figure}[htbp]
\includegraphics[width=2in,clip]{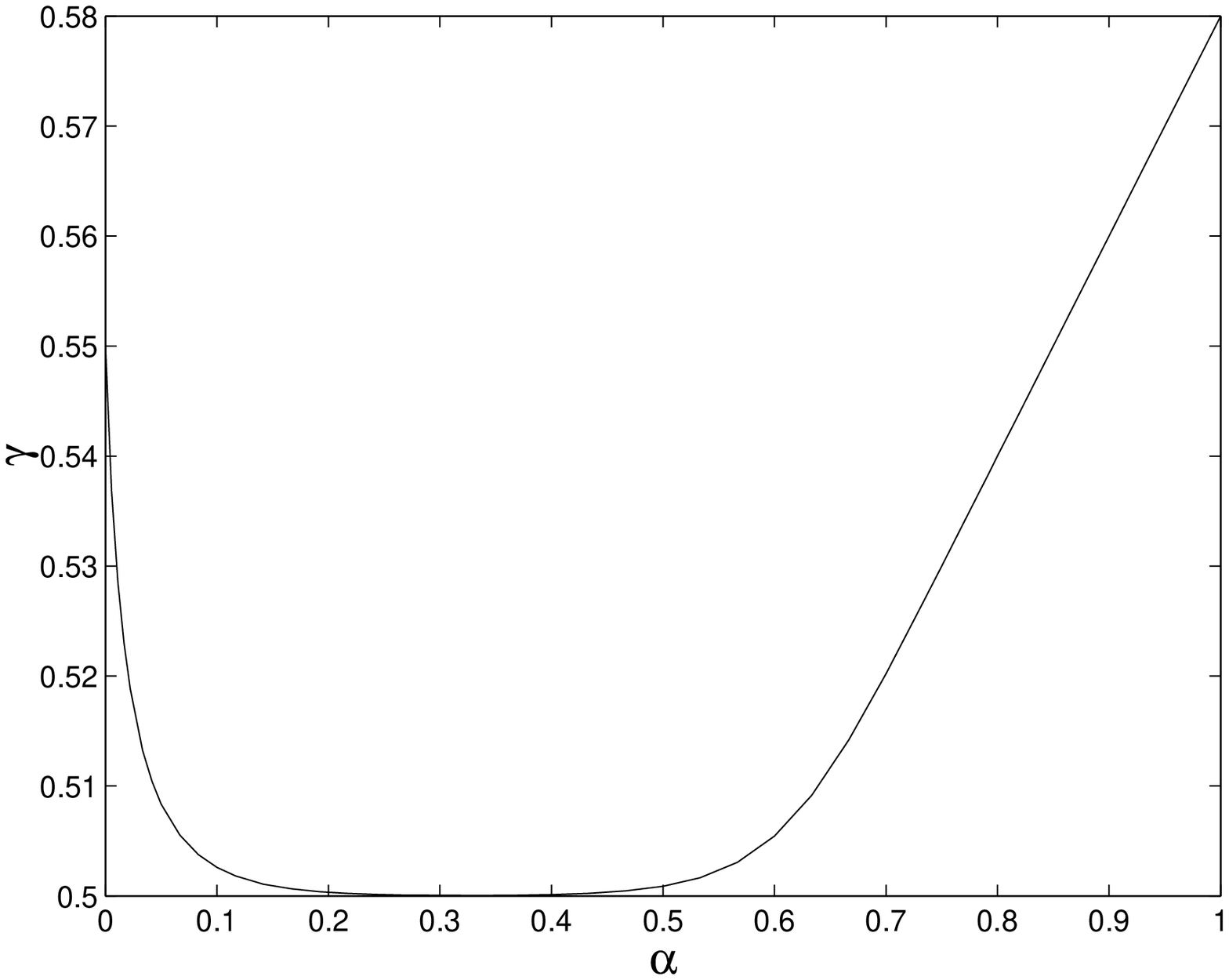}
\caption{Density profile in the  high-density maximal current (HM) phase
with $\alpha=.55$, $\gamma=.58$ $\epsilon=.001$ and $\Omega=.2$.}
\label{fig:hm}
\end{figure}
}
\newcommand{\figcontours}{%
\begin{figure}[htbp]
\includegraphics[width=3.0in,clip]{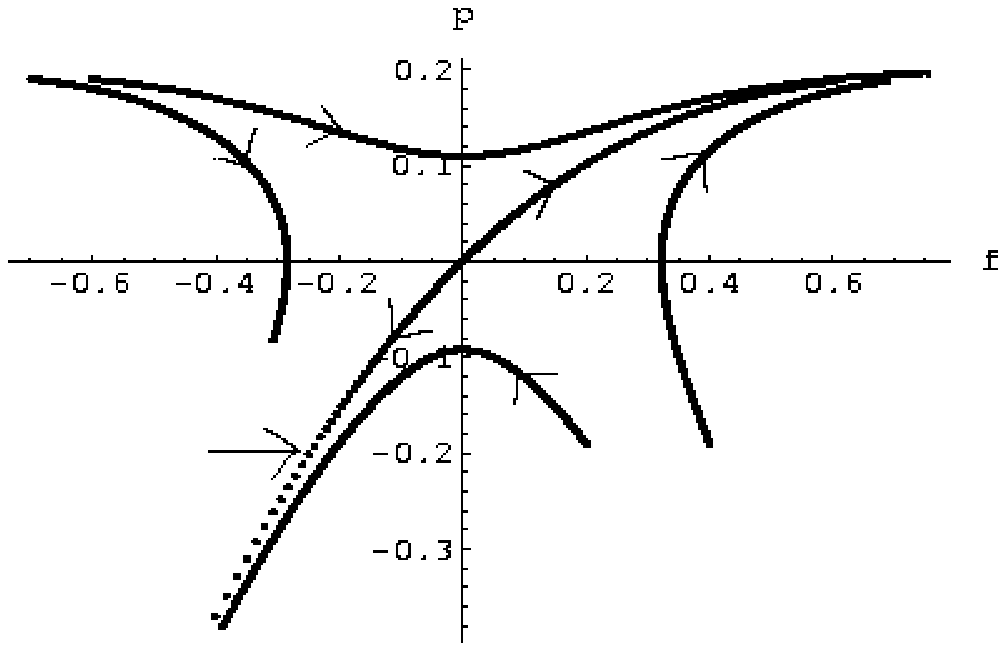}
\caption{Contour plots from equations (\ref{feq}) and (\ref{peq}) 
with $\Omega=.1$. The arrows on different 
contours indicate directions of increasing $x^*$.}
\label{fig:contours}
\end{figure}
}
\newcommand{\figlmlk}{%
\begin{figure}[htbp]
\includegraphics[width=3.0in,clip]{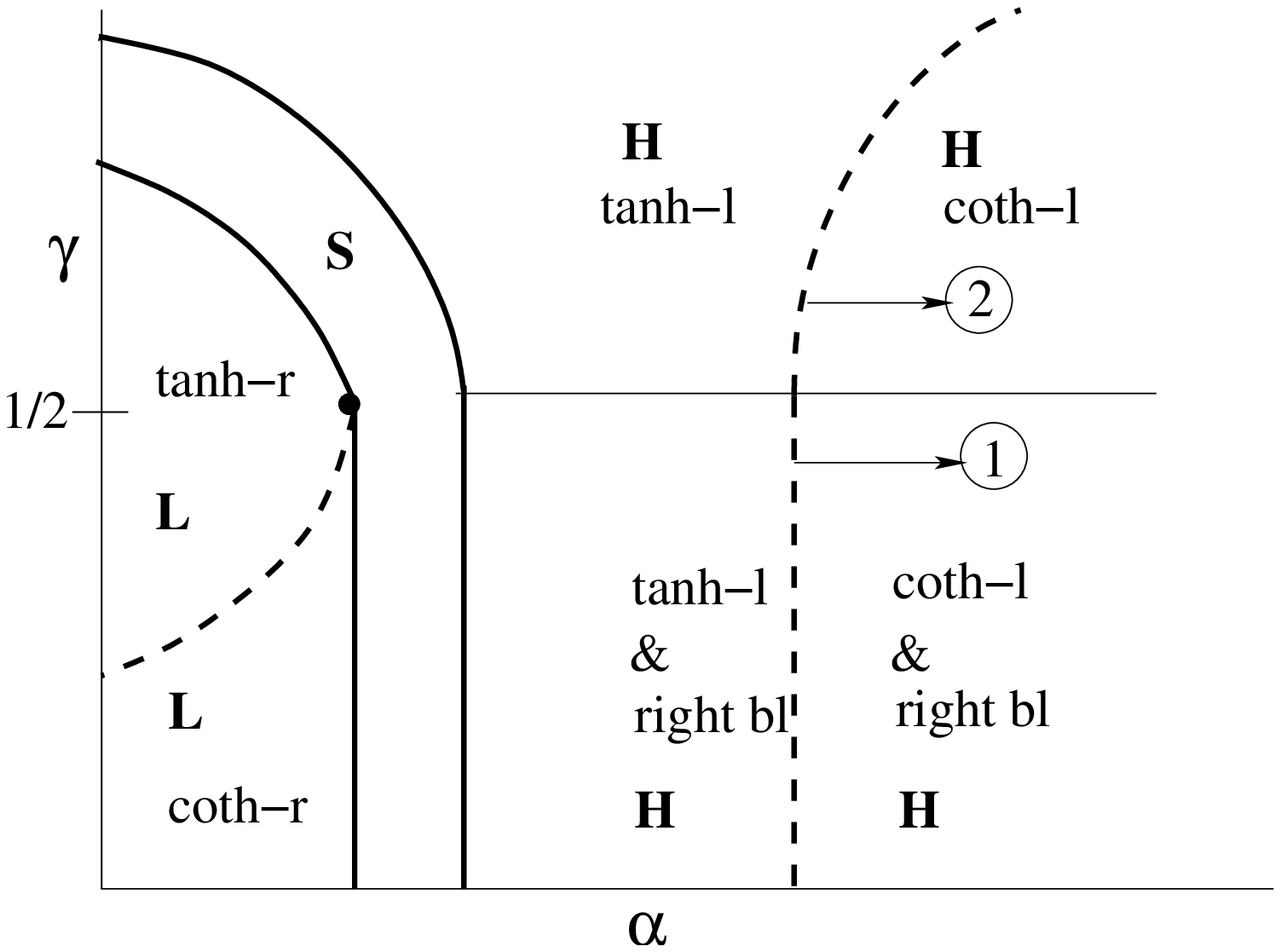}
\caption{ Phase diagram for $K\neq 1$ and $\Omega=\omega_d N<.5$.
$N$ represents the number of lattice points. 
Dashed lines represent surface transition lines. "bl" implies
boundary layer. "l" and "r" indicate presence of boundary layers 
at  left and right boundaries respectively. The filled black circle 
represents the critical point.}
\label{fig:lmlk}
\end{figure}
}
\begin{document}
\title{\bf{Bulk and surface transitions  in 
asymmetric simple exclusion process: Impact on boundary layers}}
\author{Sutapa Mukherji and Vivek Mishra}
\affiliation{Department of Physics, Indian Institute of Technology, 
Kanpur 208016, India }
\date{\today}
\begin{abstract}
In this paper, we study  boundary-induced  phase transitions 
in a particle non-conserving asymmetric simple exclusion 
process with open boundaries. 
Using   boundary layer analysis, we show 
that the key signatures of various bulk  phase transitions
are present in the boundary layers of the  
density profiles. In addition, 
we also find possibilities  of  surface transitions in the low- and high- 
density phases. The surface transition in the low-density phase 
provides a more complete description of the non-equilibrium 
critical point found  in this system.

\end{abstract}
\maketitle

\section{introduction}
Studies on a certain class of non-equilibrium systems, namely driven
diffusive systems, have revealed many features that are unexpected in
systems in thermal equilibrium \cite{schuetzrev,ligett}. 
For example, it is well known 
that although  in thermal equilibrium, one dimensional
systems with short-range interaction  cannot exhibit phase transition
 or spontaneous symmetry breaking,
 certain driven diffusive systems do exhibit such phenomena even in
one dimension \cite{evans1}. 
Asymmetric simple exclusion processes (ASEP), which 
involve biased hopping  of particles  in one direction
 with hard-core exclusion along a one-dimensional 
lattice,  fall in the class of driven diffusive systems and 
despite  its simplicity, such processes 
 are capable  of exhibiting variety of phenomena
\cite{derrida1,goutam,foster}. 
Violation of detailed balance due to the presence of a finite particle 
current  makes
such systems so special. This, for example,  is reflected in the 
boundary 
induced phase transitions in ASEP on  one-dimensional 
finite lattice \cite{krug},
where, unlike equilibrium, the  effect of boundaries 
propagates into the bulk by a  finite  particle 
current.

In this paper, we consider a particle non-conserving version of ASEP with 
open boundaries.
Particles, that are  injected at one boundary with  a rate $\alpha$,
 hop forward with rate unity (no backward hopping is allowed) 
 with hard-core exclusion  till they reach the other 
 end of the lattice where particles are withdrawn at a rate 
$1-\gamma$ \cite{frey,klumpp}.
In addition, there is  particle non-conservation at the bulk  
due to the possibility of 
attachment (detachment) of the particles to (from) the 
lattice with rate $\omega_a$($\omega_d$).  
With the variation of the injection and withdrawal rates
at the boundaries, 
such systems exhibit different phases which are
characterized by the shape of the density profiles and the nature of
the current densities\cite{frey}. A mean-field analysis,  exact 
for the hard-core case, leads to a  rich phase diagram  
(see figure \ref{fig:lml} and \ref{fig:lmlk})   compared 
to systems without particle adsorption$/$desorption  
kinetics (Langmuir Kinetics). 
For example,
there is a phase where the  density profile exhibits a 
jump from  a low to a high value at some point in the bulk.
This new phase  known as the shock phase(S)  appears in addition to  
the high(H) and low-density(L) phases where the bulk-density, 
though not constant,
remains above and below half respectively. 
Although the shock phase is present for both 
 $K=1$ and $K\neq 1$, with  $K=\omega_a/\omega_d$, these two cases  
are distinctly different since in the first case
a special particle-hole symmetry 
allows a  maximal current(M) phase  with bulk density 
equal to $1/2$. M phase and the coexistence of M phase with L and$/$or 
H phase (LMH, LM and HM phases) for $K=1$, differences in the 
 shapes of the phase boundaries
and in the nature of the 
phase transitions in the two cases make the problem 
altogether nontrivial.

In a recent work \cite{smb}, Mukherji and  Bhattachrjee
have studied the phase-transition between  low-density and  shock 
phases using boundary layer techniques. 
 Although the phase diagram 
has been studied numerically or  through a direct solution of 
 the mean-field equation earlier, the
boundary layer analysis in \cite{smb} reveals a new feature 
that the transition to the S phase from the L phase 
has a precursor of a critical deconfinement of a boundary layer near the 
open end. 
In addition, this approach provides a general 
framework for  characterizing the transitions
 for different values of $K$ and also  for studying 
different mean-field models
 with additional  inter-particle interactions.
The purpose of the present work is to employ the  boundary layer theory 
to the mean field version of the model 
to understand various  phase transitions 
from a boundary layer point of view. 
From this analysis, it becomes clear   that the 
divergences of various length scales associated 
with the boundary layers indicate the emergence of bulk phase 
transitions. Our analysis also reveals possibilities of 
 surface  transitions  
within the low- and high-density phases for both $K=1$ and $K \neq 1$.   
Except for the analysis of the surface transition, the discussion 
is primarily limited to $K=1$ since, for $K=1$, 
 various phases and  phase transitions 
give more scope for studying the effect of bulk phase transitions on 
the boundary layers.

The paper is  organized as follows. 
The model and its symmetry properties 
have been discussed in section II.
We discuss the boundary layer analysis for this model in  section III.
This section is divided into three subsections for the discussion of 
low-density, high-density, and 
maximal current related phases.
Observations from the above analysis related to the behavior 
of the boundary layers near the phase boundaries are mentioned in section 
IV. 
Finally, we conclude with a summary of our results in section V.  
\figlml
\figlmlk

\section{Model}
To describe the ASEP of noninteracting single species of particles, we
consider an one dimensional chain of $N$ lattice points and length $l$. 
Denoting the 
 occupancy of the $i$th site by  $\tau_i$, which assumes values $0$
or $1$ depending on whether the site is empty or occupied by a
particle, one may write down the mean-field  master equation describing the 
 time evolution of 
$n_i=\langle \tau_i\rangle$, 
with $<...>$ denoting the statistical average, as
\begin{eqnarray}
\frac{dn_i}{dt}=n_{i-1}(1-n_i)-
n_i(1-n_{i+1})+\nonumber\\
\omega_a(1-n_i)-\omega_d n_i.
\end{eqnarray}
In this equation, the mean-field approximation is implemented by neglecting 
correlation as 
\begin{eqnarray}
\langle \tau_i\tau_{i+1}\rangle=\langle\tau_i\rangle\langle\tau_{i+1}\rangle.
\end{eqnarray}
 In the large $N$ limit, with 
the lattice spacing $l/N\rightarrow 0$,
one can go over to the continuum by substituting 
$
<\tau_{i\pm 1}>=\rho(x,t)\pm \frac{1}{N} \frac{\partial \rho}{\partial
x}+\frac{1}{2 N^2}\frac{\partial^2\rho}{\partial x^2}....\, 
$
where  $\rho(x,t)$, is an average density  
at position $x=i\ l/N$. The choice $l=1$ simplifies the notation and 
restricts the variable $x$ within a range $\{0,1\}$. 
Keeping terms up to $O(N^{-2})$, one obtains the following equation
describing the shape of the density profile in the steady state
\begin{eqnarray}
{\epsilon}\frac{d}{dx} f_2(\rho)\frac{d\rho}{d x}+
f_1(\rho) \frac{d\rho}{dx}+\Omega f_0(\rho)=0,\label{main1}
\end{eqnarray}
with 
$\Omega=\omega_d N$, $\epsilon=1/(2N)$  and $f_i(\rho)$ for  
$i=0,\ 1,\ 2$ given as follows.  
For the dynamics that we have considered here,  
\begin{eqnarray}
f_2(\rho)=1, \ \ f_1(\rho)=2\rho-1, \ \ f_0(\rho)=K(1-\rho)-\rho. \label{ffun}
\end{eqnarray}
These $f$ functions, in general, contain
information about the dynamics. Such general form for equation
(\ref{main1}) will be useful laetr to make certain general predictions 
about the boundary layers. 

The last term in equation (\ref{main1}) originates from particle 
adsorption/desorption kinetics and  is responsible for the loss of 
particle number conservation in the bulk. In the absence of this
term, the full equation describing the time evolution of $\rho(x,t)$,
is expressible as a continuity equation 
\begin{eqnarray}
\frac{\partial\rho(x,t)}{\partial t}=-\frac{\partial j}{\partial x},
\end{eqnarray}
with the particle current-density $j$ given as 
\begin{eqnarray}
j=-\epsilon f_2(\rho) \frac{\partial\rho}{\partial x}-\hat{f_1}(\rho),
\quad {\rm where} \quad \frac{\partial \hat{f_i}(\rho)}{\partial \rho}
=f_i(\rho).
\end{eqnarray}
In  case of equation (\ref{ffun}), the current is given by 
\begin{eqnarray}
j=-\epsilon \frac{\partial\rho}{\partial x}+\rho(1-\rho).
\end{eqnarray} 
In the coninuum limit ($\epsilon\rightarrow 0$), the current-density 
$j=\rho(1-\rho)$ is bounded, $j\le 1/4$. Since, 
in the maximal current phase, the bulk density 
is $\rho=1/2$,  the current-density acquires
its maximum value $j=1/4$ in this phase.


\subsection{symmetries}

A better understanding of the  phase diagram  can be 
gained from the particle-hole symmetry of the problem. The hopping 
of a particle  in the forward 
direction is equivalent to the hopping of a hole in the 
backward direction. Similarly, the  injection of  particles at one 
boundary with a certain rate is 
equivalent to the withdrawal of holes with the same rate. The attachment
 or detachment of particles 
 can be interpreted as detachment or attachment of holes.
The invariance of 
equation (\ref{main1}) along with f-functions in equation  (\ref{ffun})
under the transformation 
\begin{eqnarray}
\rho \rightarrow 1-\rho, \ \ \
x\rightarrow 1-x\\
\omega_a \leftrightarrow\omega_d\ \ \
\alpha\leftrightarrow 1-\gamma.
\end{eqnarray}
implies that the particle-hole symmetry is respected by the system. This
symmetry is not necessarily an obvious property of the system and can be 
 easily destroyed by additional symmetry breaking interaction terms 
\cite{krug,smb}.

The situation with $K=1$ is somewhat special.  If the adsorption/desorption
kinetics were the only dynamics, the system would have settled 
in a steady-state density, known as the Langmuir density
\begin{eqnarray}
\rho_l=K/(K+1)
\end{eqnarray}
 determined from $f_0(\rho)=0$. 
If the hopping rules respect conservation, then in the steady state there
should be a homogeneous current, if this is the only dynamics.  The
corresponding density $\rho_c$ is determined from the zero of $f_1(\rho)$,
i.e.  
\begin{eqnarray}
f_1(\rho_c)=0.
\end{eqnarray}
In case of particle-hole symmetry, we expect $\rho_c=1-\rho_c$, i.e.,
$\rho_c=1/2$ to be special.  
This as noted earlier is the maximal current
state.
If $\rho_c=\rho_l$, then $\rho=\rho_c$ becomes a particular solution 
of the equation. $\rho_c$ is the density at which  bulk may allow 
nonanalytic behavior in the density.  This feature is useful for
 shock formation \cite{smb}
 though the discontinuity is  rounded by $\epsilon$ dependent 
term in equation (\ref{main1}). 
 The adsorption/desorption dynamics 
need not respect this symmetry of hopping and hence the two densities 
need not be equal. 
$K=1$ is a special case where the two densities become 
equal and the bulk dynamics is symmetric under the transformation
$\rho(x)\rightarrow 1-\rho(1-x)$.

\section{Boundary layer analysis}
By changing the boundary values $\alpha$ and $\gamma$, one may map out
all  possible steady state configurations, thereby constituting a
``phase diagram'' of the non-equilibrium system in the
$N\rightarrow\infty$ limit. To understand all these phases, we use a 
leading order boundary layer analysis that provides a systematic 
way to generate an uniform approximation of the solution 
of equation (\ref{main1}).
 In all the phases, 
the density profile over almost the  entire space, 
is described by the solution of the first order equation 
obtained by ignoring 
the second derivative term ($\epsilon\rightarrow 0$) 
in equation (\ref{main1}). 
This solution, known as the outer solution, 
 is not, in general,  expected to satisfy both 
the boundary conditions. 
In order to satisfy the boundary conditions appropriately, 
there appear special regions with  boundary layers or shocks. 
Description of these special regions requires 
going beyond the first order 
equation. The solutions describing the boundary layers or shocks
 are known as inner solutions. The constants in different solutions 
of the differential equations are determined either by 
 the boundary conditions or by smooth joining of the inner and 
outer solutions.

The low and high-density phases are related to each other due to 
particle-hole symmetry and does not require any separate treatment. 
The difference between the low-density and the maximal current phase 
arises from the outer solution itself.  
The outer solutions, solutions of  the first order equation 
(equation (\ref{main1}) in the limit $\epsilon\rightarrow 0)$, are  
\begin{eqnarray}
\rho_{1,\rm out}=1/2,\ \ \ \ \rho_{2,\rm out}=\Omega x+c, 
\ \ {\rm for}\ K=1\label{k1out}\label{out1}\\
{\rm and}\ \ 
\Omega x=g(\rho_{\rm out})-c, \ \ \ \ {\rm with}\label{out2}\\
g(\rho)=\frac{1}{1+K}(2\rho+\frac{K-1}{K+1} \log[K-(1+K)\rho])\label{grho}\\ 
{\rm for} \ \ K\neq 1.\nonumber
\end{eqnarray}
Here, $c$ is an unknown constant to be determined from the boundary condition.
In the low- and high-density phases, the density profile, over almost the 
entire space, is described 
by the linear solution  in equation (\ref{out1}) for $K=1$ or by 
the solution in equation (\ref{grho}) for $K\neq 1$. The maximal current 
phase, present only for $K=1$, has a constant density profile 
described by the  outer solution $\rho_{1,\rm out}$. 
In the phase diagram for  $K=1$, there are other regions with coexistence 
of the maximal current phase with low- or high-density phases or both.   
The density profile in these phases 
has constant part $\rho_{1,\rm out}(x)=1/2$
as well as linear parts $\rho_{2,\rm out}=\Omega x+c$ with different parts 
joined smoothly through specific inner solutions. 

The scheme to find the inner solution varies depending upon the kind of 
matching conditions the inner solution  has to satisfy. The procedure to 
find out the inner solution for L/H phases  is different
from that for phases involving M phase. The difference arises
because in the later case, the  inner solutions are  required to saturate
 to $\rho(x)=1/2$ at either of the two sides and  at this value of $\rho$,
$f_1(\rho)=0$.  In the low- or high-density phases, the boundary layer
is not required to saturate to $1/2$. 
Because of these differences, we discuss , L, H and M-related phases in
different subsections.

\subsection{Low-density phase: Surface transition}

In the low-density phase, $c=\alpha$ for $K=1$ and $c=g(\alpha)$ for 
$K\neq 1$ since 
 the outer solution satisfies the left boundary condition. To find the 
inner solution, one needs to express equation (\ref{main1}) in terms
of $\tilde x=(x-x_d)/\epsilon$, where $x_d$ denotes the location of the 
solution. In the $\epsilon\rightarrow 0$ limit, the inner solution 
is the solution of 
\begin{eqnarray} 
\frac{d\rho_{\rm in}}{d\tilde x}=\frac{F(\rho_{\rm in})}{f_2(\rho_{\rm in})},
\ \  \ {\rm where}\label{inner1}
\end{eqnarray}
\begin{eqnarray}
F(\rho)\equiv \hat f_1(\rho_o)-\hat f_1(\rho).\label{bigf}
\end{eqnarray} 
The matching condition  
$\rho_{\rm in}(\tilde x\rightarrow -\infty)=
\rho_o\equiv \rho_{2,{\rm out}}(x=1)$ for smooth joining of two solutions 
has already been incorporated. The inner region obeys the particle 
conservation condition and the continuity equation demands a homogeneous 
current. Consequently, the current has to be equal to the bulk current 
entering the region. This is the content of equations (\ref{inner1}) and 
 (\ref{bigf}) though 
derived in a different way.

The condition for saturation of the 
inner solution as $\tilde x\rightarrow \infty$ requires 
\begin{eqnarray}
F(\rho)=0 \ \ {\rm for}\ \rho=\rho_s>\rho_o.
\end{eqnarray}  
If $\gamma>\rho_{\rm in}(\tilde x\rightarrow \infty)$, 
the boundary  condition cannot be satisfied by $\rho_{\rm in}$.
As a result, the surface layer deconfines from the surface 
and enters into the bulk with the outer solution again appearing at the right 
edge to satisfy the right boundary condition. This mechanism leads to the 
formation of a  shock in the density profile at the bulk. The bulk 
transition  to the shock phase, therefore,
 has a precursor of deconfinement of the 
surface layer. Since 
$\rho_o$ is a function of $\alpha$, one has a phase boundary 
 on the $\alpha-\gamma$ plane 
$\gamma=\rho_s(\rho_o(\alpha))$ that separates the low-density phase 
from the shock phase. From equations (\ref{ffun}) and (\ref{bigf}), 
we find $\rho_s=1-\rho_o$.  Assuming simple zeros for $F(\rho)$, 
we write 
\begin{eqnarray}
F(\rho)=-(\rho-\rho_o)(\rho-\rho_s)\phi(\rho),\label{Ffactor}
\end{eqnarray}
where $\phi(\rho)$ can be determined  using equation (\ref{main1}). 
This form of $F(\rho)$ is convenient as it  yields the length scale 
associated with the crossover of the surface profile 
to the bulk profile quite generally. The large $\tilde x$ behavior of the
inner solution can be found from  
\begin{eqnarray}
\frac{d\rho_{\rm in}}{d\tilde x}\sim -\frac{(\rho-\rho_s)}{w(\alpha)},
\label{lengthscale}
\end{eqnarray}
where 
\begin{eqnarray}
w(\alpha)=(\rho_s-\rho_o)^{-1} \frac{f_2(\rho_s)}{\phi(\rho_s)}.
\end{eqnarray}
Equation (\ref{lengthscale}) shows $w(\alpha)$ as the characteristic 
length scale for the approach to saturation or to the bulk density.
 $w(\alpha)$ can be made to diverge by changing $\alpha$ and this 
locates a  critical point on the phase boundary between the 
low-density and the shock phase. At this critical point 
$(\alpha_c,\gamma_c)$, $\rho_s=\rho_o=\gamma_c$. 

Given this form of $F(\rho)$, in the low-density phase, 
the slope of the inner solution  
  at $x=1$, i.e. at $\rho=\gamma$,
   is positive if $\rho_o<\gamma<\rho_s$ 
and negative if $\gamma<\rho_o$. Thus, with $\gamma_{\rm sc}=\rho_o$, 
there is an increasing 
solution at $x=1$ for $\gamma>\gamma_{\rm sc}$ and a decreasing one
for $\gamma<\gamma_{\rm sc}$. This leads to a general conclusion, 
 that for every $\alpha$, if there is a bulk transition at 
$\gamma=\rho_s(\rho_o(\alpha))$, there is a "boundary" transition 
at $\gamma=\rho_o(\alpha)$.  This defines a surface transition line
$\gamma=\gamma_{\rm surf}(\alpha)$ in the low-density phase.  
The transition is strictly at the boundary 
because as one crosses the phase boundary, the bulk density profile remains 
same but the boundary profile changes drastically.    Such  surface transition 
 is expected to be true generally for all $K$ and also for 
interacting system, wherever a transition to a shock phase takes place.

In terms of   $w$ specified above, 
the inner solution, in general,  has a form 
\begin{eqnarray}
\rho_{\rm in}(\tilde x)=\rho_{\rm o} S_{\rm in}(\tilde x/2 w+\xi), 
\end{eqnarray}
with  $S_{\rm in}\rightarrow 1$ as $\tilde x \rightarrow -\infty$. 
The constant, $\xi$, can be determined from the constraint 
$S_{\rm in}(\tilde x=0)=\gamma$ and the explicit form of 
$F(\rho)$ specified in  equation (\ref{Ffactor}). Near the surface 
transition line $\xi$ has a  logarithmic divergence 
\begin{eqnarray}
\xi \sim \ln\mid\gamma-\rho_o\mid,
\end{eqnarray}
which  leads us to define an  exponent $\zeta_{\rm s}$ as 
\begin{eqnarray}
\xi \sim \mid \Delta\alpha\mid^{\zeta_s},\ \  {\rm for}\ \ 
 \alpha=\alpha_{\rm surf}\pm \Delta\alpha.
\end{eqnarray}
In our case, 
\begin{eqnarray}
\zeta_{\rm s}=O(\log).
\end{eqnarray}

In the following subsection, 
we illustrate this general approach for special cases of $K=1$ and $K\neq 1$.

\subsubsection{Results for $K=1$ and $K\neq 1$}

Following the above arguments,
 surface transition lines can be obtained for all values of $K$.
For all $K$, the  explicit inner solution with positive slope at $x=1$  
is \begin{eqnarray}
\rho_{\rm in}(\tilde x)
=\frac{1}{2}+\frac{(1-2\rho_o)}{2}\tanh[\frac{\tilde x}{2w}+\xi],\label{tanh}
\end{eqnarray}
where  $\xi$ is a  constant and 
\begin{eqnarray}
w=1/(1-2 \rho_o).\label{wgen}
\end{eqnarray} This   solution appears for $\gamma>\rho_o$. 

As $\tilde x\rightarrow \infty$, the inner solution saturates to  
$\rho_s=1-\rho_o$. The bulk transition to the shock phase occurs when the 
 saturation valuei, $\rho_s$, of the surface layer, is smaller 
than $\gamma$. The 
phase boundary between the low-density and the shock-phase is, therefore,
given by the equation $1-\rho_o(\alpha)=\gamma$. 
The surface transition to a boundary layer of negative 
slope occurs when  $\gamma<\gamma_{\rm sc}=\rho_o$.  The boundary 
layer, in this case is 
\begin{eqnarray}
\rho=\frac{1}{2}+\frac{1-2\rho_o}{2}\coth[(\frac{\tilde x}{2w}+\xi)],
\label{coth}
\end{eqnarray}
where $w$ is same as in equation (\ref{wgen}).   
$w$ and $\xi$ together determine the position of the  "virtual origin" at 
which the argument of the inner solution vanishes. This is simply in the 
sense of mathematical continuation since the virtual origin 
 may lie well beyond 
the physical range of $x$, $\{0,1\}$, with an unphysical value of the 
density. 
\figtanh
\figcoth
 
For $K=1$, using 
$\rho_o=\Omega+\alpha$, we find 
\begin{eqnarray}
w=1/(1-2\Omega-2\alpha).
\end{eqnarray}
Further, the  constraint  $\rho_{\rm in}(\tilde x=0)=\gamma$, leads to   
\begin{eqnarray}
\xi=\frac{1}{2}
\log[\frac{\gamma-\Omega-\alpha}{1-\Omega-\alpha-\gamma}].\label{tanhlow}
\end{eqnarray}
for $\tanh$ type surface layer (see figure \ref{fig:tanh}) and 
\begin{eqnarray}
\xi=\frac{1}{2}
\log[\frac{\gamma-(\Omega+\alpha)}{\gamma-(1-\Omega-\alpha)}]
\end{eqnarray} for $\coth$ boundary layer (see figure \ref{fig:coth}).
The surface transition from the $\tanh$ type surface layer 
 to the  $\coth$ type 
boundary layer takes 
place if $\gamma<\gamma_{\rm sc}=\rho_{\rm o}=\Omega+\alpha$ 
leading to a linear 
surface transition line, $\gamma=\Omega+\alpha$,  on the 
$\alpha-\gamma$ plane. As this line is approached from either of the two 
low-density phases, $\xi$ diverges logarithmically as
\begin{eqnarray}
\xi \sim \log\mid \gamma-\Omega-\alpha\mid.
\end{eqnarray}
The surface transition line 
$\rho_o(\alpha)=\gamma$ and the shock phase boundary 
$\rho_{\rm s}=1-\rho_o(\alpha)=\gamma$ intersect at $\rho_{\rm o}=1/2$. 
This intersection point is, therefore, also the critical point 
$(\alpha_c,\gamma_c)$ at which $w(\alpha)$
diverges.  For $K=1$, the critical point is $(1/2-\Omega,1/2)$. 
For   $\gamma<\gamma_{\rm c}$, this boundary layer 
with negative slope at $x=1$ cannot shocken if $\alpha$ is increased. 
However, as $\alpha$ is increased, this decaying boundary layer 
continues to be there in the  low-density 
maximal-current (LM) phase.

For $K\neq 1$, $\rho_o$ is the   solution of $g(\rho_o)=\Omega+g(\alpha)$ 
with $g(\rho)$ given in 
equation (\ref{grho}). In this case, therefore,  the 
surface transition line $\gamma=\gamma_{\rm sc}=\rho_o$
 is determined from the solution for $\gamma$  from the equation 
\begin{eqnarray}
g(\gamma)=\Omega+g(\alpha).\label{surfacetrans}
\end{eqnarray}
The condition for shock formation from a low-density phase with $\tanh$ type 
boundary layer, 
on the other hand,  leads to a 
phase boundary given by the  solution of the equation
\begin{eqnarray}
g(1-\gamma)=\Omega+g(\alpha).\label{shockboundary}
\end{eqnarray}
\figintersect
The solutions for $\gamma$  from  equations (\ref{surfacetrans}) and 
(\ref{shockboundary}) are symmetric around $\gamma=1/2$ 
(see figure \ref{fig:intersect}) and  
the intersection of the two solutions at $\gamma=\gamma_c=1/2$ 
 is the critical point 
$(\alpha_c,\gamma_c)$ where 
\begin{eqnarray}
g(\gamma_c)=\Omega+g(\alpha_c).
\end{eqnarray}
Since the surface transition line is symmetric to the phase boundary 
between  
the shock and  the low-density phases,  shapes of these
to lines close the critical point are same \cite{smb}.
The shape of the surface transition line can be obtained
independently by substituting  
$\gamma=\gamma_c-\Delta \gamma$ and $\alpha=\alpha_c-\Delta\alpha$ in  
equation (\ref{surfacetrans}) and  expanding it  in small $\Delta \gamma$ and 
$\Delta\alpha$. This leads to a general equation 
\begin{eqnarray}
g^{\prime}(\gamma_c) \Delta \gamma-g^{\prime\prime}(\gamma_c) 
(\Delta\gamma)^2/2+....=g^{\prime}(\alpha_c) \Delta\alpha.
\end{eqnarray}
Since for $K\neq 1$, $g^\prime(\gamma_c)=0$, the shape of the surface 
transition line near the critical point is given by   
$\Delta\gamma \sim \Delta\alpha^{\chi_-^{\rm s}}$ with 
\begin{eqnarray}
\chi_-^{\rm s}=1/2.
\end{eqnarray}
Similar to the $K=1$ case, the 
 $\coth$ boundary layer, in the low-density phase,  
leads way  to a  decaying boundary layer in the shock phase 
 for $\gamma<\gamma_c$. 

The mechanism for the  formation of the shock for 
$\gamma>\gamma_c$,  is different from that for $\gamma<\gamma_c$,
 since in the latter case, there is no $\tanh$ type 
boundary layer to be deconfined to form a shock. 
 The effective boundary 
condition  for shock formation, for $\gamma<\gamma_c$
 is same as $\gamma=\gamma_c$ 
since  the right branch of  
the outer solution satisfies  this effective boundary condition at 
its right edge. The original boundary condition  
$\rho(x=1)=\gamma$ with $\gamma<\gamma_c$ is satisfied 
finally with a decaying boundary layer. 
 As a consequence of this,
 $\alpha_c$ continues to be the critical value of 
$\alpha$ for shock formation for all $\gamma<\gamma_c$ leading to a
vertical phase boundary between the low-density and the shock phase.
As $\alpha$ increases in the shock phase, the discontinuity, 
formed at $x=1$, moves towards  $x=0$ till it 
reaches the other end  at the high-density-shock phase boundary. 
Whereas for $\gamma>\gamma_c$, the deconfined 
boundary layer or the shock at $x=1$ has
a finite height, for $\gamma<\gamma_c$, the shock 
  height increases  continuously from zero to a finite value 
as $\alpha$ is increased. 
Although the
mechanism of shock formation is different for $\gamma>\gamma_c$ and
$\gamma<\gamma_c$, the discontinuity is always described by a $\tanh$
type inner solution.
Alternatively,  for all values of $\gamma$, 
the emergence of shock, as one approaches the shock phase from the 
high-density side, is through the deconfinement   
of a $\tanh$ boundary layer of finite height at $x=0$. As a consequence of 
this,  the high-density-shock 
phase-boundary cannot have any critical point on it.   

The entire process of surface transition and then shock formation
by changing $\gamma$ for a given $\alpha$ can be 
understood on a more physical ground. 
 In the low-density phase,
for small $\gamma$, the withdrawal rate at $x=1$ is high. 
Since the bulk dynamics  is completely controled by hopping and 
adsorption/desorption kinetics, a large withdrwal rate with a fixed 
$\alpha$, causes 
particle depletion at $x=1$ to be described by a "virtual origin" 
somewhere at $x>1$. As $\gamma$ is increased, the withdrawal rate 
decreases and we reach a situation where there is neither any 
depletion nor any accummulation of particles at the end. 
The "virtual origin" 
is shifted to $\infty$ now. If $\gamma$ is increased further, 
the withdrawal rate is too slow to get rid of particles that 
bulk dynamics can feed. This leads to an accummulated region 
    at the boundary until a stage  where this accummulated region 
becomes macroscopic. Beyond this point, the withdrawal rate controls 
the density near the end and the injection rate controls the remaining
part of the density with the two parts joined through a shock or 
dicontinuity.

\subsection{High-density phase}
\subsubsection{$K=1$}
For $K=1$, 
the H-phase can be completely understood from the knowledge in L-phase 
by exploiting the particle-hole symmetry. In this sense, 
the high-density phase for particles is equivalent 
 to the low-density phase 
for holes with density profile same as that in figure \ref{fig:tanh} or 
figure \ref{fig:coth}
when the coordinates are transformed   as $x\to 1-x$.
In this  transformed coordinates, above the line $\gamma=\Omega+\alpha$, 
 we have the density profile for  
holes same as figure \ref{fig:tanh}. The boundary layer here is 
 described by equation (\ref{tanh}) with the constants 
\begin{eqnarray}
w=1/(2\gamma-2\Omega-1) \ \ {\rm and}\label{bhigh1}\\
\xi=\frac{1}{2}
\log[(\gamma-\Omega-\alpha)/(\gamma+\alpha-1-\Omega)].\label{bhigh}
\end{eqnarray}
\figtanhhigh
It is straightforward to verify that  in original coordinates, the particle 
density profile appears as  figure \ref{fig:tanh-high}.
It consists of a linear profile satisfying the right 
boundary condition and  a boundary layer as in equation 
(\ref{tanh}) with 
\begin{eqnarray}
\xi=\frac{1}{2}
\log[(\gamma+\alpha-1-\Omega)/(\gamma-\Omega-\alpha)],
\end{eqnarray}
and $w$ same as in (\ref{bhigh1}). 
The particle-hole symmetry also guarantees a surface phase transition in 
the high-density phase across $\gamma=\Omega+\alpha$ line. 
\figcothhigh
The  $\tanh$ type boundary layer at $x=0$ with positive slope 
 changes to  a
boundary layer with negative slope (see figure \ref{fig:coth-high})
described by equation (\ref{coth}) with constants that can be determined 
using the symmetry.

\subsubsection{$K\neq 1$}
The high-density phase can be divided into two major parts.
For $\gamma<\gamma_c=1/2$,
the density profile has boundary layers on both the ends. In addition to a
$\tanh$ type boundary layer at $x=0$, there is  a decaying boundary layer
at $x=1$. The right boundary layer helps the density profile satisfy
the boundary condition at $x=1$ from a value
$\rho=1/2$. Thus as in the case of shock phase with $\gamma<1/2$,  the
effective boundary condition on the right edge for the outer solution
continues to be $\rho=1/2$ in this part of the H phase.
In the other part of the H phase, the density profile has only one
$\tanh$ boundary layer at $x=0$.

The transition to the shock phase from the high-density phase takes 
place through the deconfinement of the inner solution 
\begin{eqnarray}
\rho=1/2+\frac{(2\rho_o'-1)}{2} \tanh[\frac{(2\rho_o'-1)\tilde x}{2}+\xi],
\label{innerhigh}
\end{eqnarray}
at $x=0$. Here  $\rho_o'=\rho_{2,\rm out}(x=0)$ is the value 
of the outer soultion 
at $x=0$. For $\gamma>1/2$, the outer solution obeying the boundary 
condition $\rho(x=1)=\gamma$ can be obtained by solving 
\begin{eqnarray}
g(\rho_o')=-\Omega+g(\gamma).
\end{eqnarray}
For $\gamma<1/2$, the outer solution satisfies the 
effective boundary condition $\rho(x=1)=1/2$  and it is the solution
 of the equation
\begin{eqnarray}
g(\rho_o')=-\Omega+g(1/2).
\end{eqnarray}   
The boundary layer deconfines, whenever $\alpha$ is smaller than $1-\rho_o'$,
 the saturation value of the inner solution in equation (\ref{innerhigh}). 
This leads to the high-density-shock phase  boundaries,
\begin{eqnarray}
g(1-\alpha)=-\Omega+g(1/2) \ \ {\rm for}\ \ \gamma<1/2\\
g(1-\alpha)=-\Omega+g(\gamma)\ \ {\rm for}\ \ \gamma>1/2.
\end{eqnarray}
Since for $\gamma<1/2$, 
the value of the critical $\alpha$ does not depend on $\gamma$,
the  phase boundary is vertical for all $\gamma<1/2$, with a $K$ dependent 
value of $\alpha$.  

Across the surface transition line, the slope of the boundary layer 
at $x=0$ changes sign.
If the  value of  $\rho_o'$ is larger than
$\alpha$, a boundary layer with a positive slope at $x=0$ is expected. 
In the reverse situation, one expects a boundary layer with a negative 
slope at $x=0$. The transition lines are, therefore, given by 
\begin{eqnarray}
g(\alpha)=-\Omega+g(1/2)\ \ {\rm for} \ \ \gamma<1/2,\label{vertical}\\
g(\alpha)=-\Omega+g(\gamma)\ \ {\rm for}\ \ \gamma>1/2.\label{curve}
\end{eqnarray} 
The surface transition lines in equations (\ref{vertical}) and 
(\ref{curve}) are represented by dashed lines $1$ and $2$, respectively, 
in figure \ref{fig:lmlk} 
These two lines meet at $\gamma=1/2$  with 
 a value of $\alpha$ that depends on $K$. The surface transitions across 
both the lines are associated with the divergence of $\xi$ and there is 
no critical point on these lines.

\subsection{Boundary layers in LM and HM phases for $K=1$}
For $K=1$, 
 $\alpha=1/2-\Omega$ and $\gamma=1/2+\Omega$ are the 
boundaries for the  low-density and high-density phases respectively since 
 $\tanh$ or $\coth$ type boundary layers are no more 
valid on these lines.  The   LM phase and the symmetrically opposite
HM phase appear in the regimes  $\{1/2-\Omega<\alpha<1/2, 
\gamma<1/2\}$  and $\{\alpha>1/2,\ 1/2<\gamma<1/2+\Omega\}$
respectively.  
\figlm
In the LM phase (see figure \ref{fig:lm}), 
the linear profile,  $\rho(x)=\Omega x+\alpha$,
satisfies the left boundary condition and 
continues till $x_{\rm cl}=(1/2-\alpha)/\Omega$ where 
$\rho(x_{\rm cl})=1/2$. The constant profile continues till the 
other end where a boundary layer finally satisfies the boundary condition. 
\fighm
 In the HM phase (see figure \ref{fig:hm}), 
the linear profile satisfying  
the right boundary condition ends at
$x_{\rm ch}=(1-2\gamma+2\Omega)/2\Omega$ with constant density profile
for  $x<x_{\rm ch}$ and a boundary layer at $x=0$.  
  To obtain the 
boundary layer near $x=1({\rm or}\  x=0)$ for LM (or HM) phase, it 
 is useful to express equation (\ref{main1}) in terms of 
\begin{eqnarray}
f(x^*)=(2\rho-1)/\sqrt{\epsilon},
\end{eqnarray}
with $x^*=(x-x_0)/\sqrt{\epsilon}$, where $x_0$, as before, represents 
the center of the solution. In terms of $f(x^*)$, the new
equation is 
\begin{eqnarray}
\frac{1}{2} \frac{\partial^2 f}{\partial
{x^*}^2}+\frac{f}{2}\frac{\partial f}{\partial x^*}-\Omega f=0.\label{phase1}
\end{eqnarray}
A phase plane analysis  of equation
(\ref{phase1}) is useful to identify the appropriate 
inner solution that can satisfy  the
boundary conditions. Since the details of such analysis can be found 
in \cite{cole} or in other mathematics text-books, we briefly mention the 
basic principles and obtain the contour-plot (figure \ref{fig:contours}) 
in an Appendix. 
 Denoting $\frac{df}{dx^*}=p$, we have 
\begin{eqnarray}
\frac{dp}{df}=f \frac{2\Omega-p}{p}.
\end{eqnarray}
To obtain the boundary layer at $x=1$ for the LM phase,
we require the specific solution 
which  satisfies the  boundary conditions 
$
\rho(x^*)=1/2, \ \  {\rm as}\ \ \   x^*\rightarrow -\infty$ and
$\rho(x^*)=\gamma \ \ {\rm as}\ \ \  x\rightarrow 1$. 
These conditions are fulfilled by the contour marked with an arrow in figure
\ref{fig:contours}.
The origin $(f=p=0)$ is a fixed point for the differential 
equations and a simple linearization around this fixed point 
leads to $p=(2 \Omega)^{1/2} f$. 
The approach to $\rho=1/2$ as $x^*\rightarrow -\infty$ is,
therefore, exponential as 
\begin{eqnarray}
f\sim   
\exp[(2 \Omega)^{1/2}x^*].\label{meethalf}
\end{eqnarray}
Away from the fixed point $(f=p=0)$, the 
solution of the differential equation is 
$f^2\sim -2p$,
implying $f\sim 2/x^*$, as $x^*\rightarrow 0$. The boundary 
condition $\rho(x=1)=\gamma$ further leads to 
$x_0=1-\frac{2\epsilon}{2\gamma-1}$. As $x\rightarrow 1$, the density 
profile thus has an algebraic decay as 
\begin{eqnarray}
\rho(x)=\frac{1}{2}+\frac{\epsilon}{x-1+2\epsilon/(2\gamma-1)}.
\label{lmalgeb}
\end{eqnarray}
 For the HM phase, the boundary layer at the left edge 
has to be described by the specific 
solution  that satsifies the boundary condition at 
$x=0$ and approaches $0$ exponentially as $x^*\rightarrow \infty$.  
It can be shown that this solution  is the same as 
(\ref{lmalgeb}) near $x=0$,  with $x_0=\frac{2\epsilon}{1-2\alpha}$.
The matching between the linear  and the constant profile around 
$x_{cl}$ or $x_{ch}$ can also be done by choosing the appropriate 
solution from the contour plot.
In the case of LM phase, this solution 
should merge to the linear one for $x^*\rightarrow -\infty$ and,
approach $0$
exponentially as $x^*\rightarrow \infty$, 
with the same length scale as in equation (\ref{meethalf}). Similar 
analysis is suitable also for the maximal current(M) phase, that appears 
for $\alpha>1/2$ and $\gamma<1/2$.
In this phase, the density profile is fixed at $1/2$ over 
the entire lattice except for  decaying boundary layers same as those in 
LM and  HM phases at $x=1$ and $x=0$, respectively.

\section{Approach to various phase boundaries}

Our final aim is to study the boundary layers for $K=1$ 
as different phase boundaries are approached. 
As the phase boundary $\alpha=1/2-\Omega$ is approached 
from the low-density side (along path 2), $w\xi$ approaches 
a finite value $1/(2\gamma-1)$ which  diverges as 
the special point $\gamma=1/2$ is approached. 
Since at this special point, 
$\Omega+\alpha=\gamma=1/2$,
 the linear profile satisfies both  the boundary conditions and there is 
no need of any boundary layer. 
Since $\coth x\sim 1/x$ as $x\rightarrow 0$, there is an 
algebraically  decaying inner solution 
\begin{eqnarray}
\rho(x)=\frac{1}{2}+\frac{\epsilon}{x-1+2\epsilon/(2\gamma-1)},
\end{eqnarray}
near the phase boundary 
$\alpha=1/2-\Omega$ for $\gamma<\gamma_c$
 as $x\rightarrow  1-\frac{2\epsilon}{2\gamma-1}$.
This decay is exactly the same as that we find in  
the density profile in equation (\ref{lmalgeb})
 that describes the boundary layer in 
the LM phase (on the other side of the phase boundary, 
$\alpha=1/2-\Omega$). From this analysis, it becomes clear, how the 
$\coth$ boundary layer transforms  to an algebraically decaying
boundary layer appropriate for  LM phase across the phase boundary.  
As $\tilde x\rightarrow -\infty$, 
the $\coth$ inner solution in the L-phase 
approaches $\rho_o=\Omega+\alpha$ exponentially as 
\begin{eqnarray}
\rho(\tilde x)=\Omega+\alpha-(1-2\Omega-2\alpha) 
e^{2[(1-2\Omega-2\alpha)\tilde x/2+\xi]},
\end{eqnarray}
with a length scale $1/(1-2\Omega-2\alpha)$ that 
diverges as the phase boundary $\alpha=1/2-\Omega$ is appraoched. 
It is interesting to note that as the phase boundary is approached, 
the algebraic decay  of the   inner solution near $x=1$ 
is  same in both L and LM phases. The approach of the two inner solutions 
in  L and LM phases to the bulk density, on the other hand, 
 is exponential with two different length scales.  
Further,  
 $x_0$ diverges as $x_0 \sim (2\gamma-1)^{-1}$ as one approaches 
 the LM and LMH phase boundary along path (3).
Situation is
symmetric as one approaches the HM and H phase boundary.
As the phase boundary  $\alpha>1/2$ and $\gamma=1/2+\Omega$ is approached
from the H phase  along path (4),
$\xi w$ in the $\coth$ boundary layer at $x=0$ appraoches $1/(2\alpha-1)$.
The power-law decay of the boundary layer near $x=0$ as 
\begin{eqnarray}
\rho(x)=\frac{1}{2}+\frac{\epsilon}{x-2\epsilon/(1-2\alpha)}
\end{eqnarray}
is the same as that for the boundary layer in HM phase.   
As HM or LM phases are approached from M phase, 
the boundary layer at $x=1$ or at $x=0$, respectively, 
 disappears and a linear profile 
at the respective edge starts appearing. This is marked by the divergences
of $x_0$ as $(2\gamma-1)^{-1}$ or $(2\alpha-1)^{-1}$ along path (5) or (6) 
respectively.

\section{summary}
In summary, we have shown that various  bulk phase transitions in  particle 
nonconserving asymmetric 
simple exclusion processes with open boundaries are associated with 
divergences of various length scales related to the  
boundary layers of the density profile. The nature of the 
divergences are obtained by implementing the 
 techniques of boundary layer analysis on the 
mean-field equation for the steady-state density profile. 
In addition to this, we also  show that the low- and high-density 
phases exist in two different surface phases with distinctly 
different surface layers in the density profile.

\appendix
\section{Phase-plane analysis for equation (\ref{phase1})}
In terms of $p$ and $f$, second order equation  (\ref{phase1}) 
can be decomposed into two coupled  first order equations
\begin{eqnarray}
\frac{df}{dx^*}=p  \ \ \ {\rm and}\label{feq}\\
\frac{dp}{dx^*}= f(2\Omega-p).\label{peq}
\end{eqnarray}
$(f=p=0)$ is a fixed point for these equations. Linearization around this 
fixed point leads to the following matrix equation 
\begin{eqnarray}
\frac{d}{dx^*}\left(
\begin{array}{c}
\delta f\\
\delta p
\end{array}\right)=\left(
\begin{array}{c}
0 \ \ \ 1\\
2\Omega\ \ 0
\end{array}\right)\left(
\begin{array}{c}
\delta f\\
\delta p
\end{array}\right),
\end{eqnarray}
which can be solved through 
standard diagonalization scheme. The eigen vectors corresponding to the 
eigenvalues $(2\Omega)^{1/2}$ and $-(2\Omega)^{1/2}$ are $\left(
\begin{array}{c}
1\\
(2\Omega)^{1/2}
\end{array}\right)$ and $\left(
\begin{array}{c}
1\\
-(2\Omega)^{1/2}
\end{array}\right)$ respectively. 
In the following, we present the contour-plots in $f-p$ 
plane for different boundary conditions. 
\figcontours

\end{document}